# Calibrated Microwave Reflectance in Low-Temperature Scanning Tunneling Microscopy

Bareld Wit[1, a)], Georg Gramse[2], and Stefan Müllegger[1]

**Affiliations**

1. Institute of Semiconductor and Solid State Physics, Johannes Kepler University Linz, 4040 Linz, Austria
2. Biophysics Institute, Johannes Kepler University, 4020 Linz, Austria

a) Author to whom correspondence should be addressed: bareld.wit@jku.at

**Abstract**

We outline calibrated measurements of the microwave reflection coefficient from the tunnel junction of an ultra-high vacuum low temperature scanning tunneling microscope. The microwave circuit design is described in detail, including an interferometer for enhanced signal-to-noise and a demodulation scheme for lock-in detection. A quantitative, *in-situ* procedure for impedance calibration based on the numerical 3-error-term model is presented. Our procedure exploits the response of the microwave reflection signal due to the change of the tunneling conductance caused by sub-nm variation of the tunneling distance. Experimental calibration is achieved by a least-squares numerical fit of simultaneously measured conductance and microwave reflection retraction curves at finite conductance. Our method paves the way for nanoscale microscopy and spectroscopy of dielectric surface properties at GHz frequencies and cryogenic temperatures. This opens a promising pathway even for dielectric fingerprinting at the single molecule limit.

**Introduction**

Recent developments of near-field scanning microwave microscopy[1] have enabled the characterization of materials with nanoscale dimensions, surfaces, and nanoelectronics devices[2] up to GHz frequencies and with sub-micron lateral resolution. In particular, this includes measuring electrostatic force[3,4], electrical capacitance[5,6] and electrochemical activity[7]. Each of these techniques has its specific limitations, such as the frequency of operation[6,8,9]. Recently, microwave reflection spectroscopy with sub-micron lateral resolution has been demonstrated up to 19 GHz[10,11]. Various material properties, including relative dielectric permittivity[12] and dopant concentration[2] have been successfully detected by a transmission line implementation realized in an atomic force microscope operated at ambient conditions. Although closely related to the atomic force microscope, the scanning tunneling microscope (STM) has been scarcely demonstrated as platform for calibrated microwave reflection spectroscopy to date. The present work outlines a straightforward approach, *i.e.* utilizing the STM tip as a probe for calibrated microwave reflection spectroscopy. In particular, we employ a low-temperature STM instrument, which is well-known to provide ultimate stability and spatial resolution[13] by operating at ultra-high vacuum conditions and cryogenic temperatures. Our approach offers great potential for achieving microwave impedance spectroscopic fingerprinting even at the scale of single molecules – thus complementing related methods that utilize spin-resonance[14,15] or dielectric relaxation[16] at GHz frequencies. A major advantage of STM over similar AFM based techniques is the ability to select the bias and current setpoint at which the microwave reflection spectroscopy is performed. This opens the door to, for example, studying transient molecular redox reactions[7].

Commonly, the tunnel junction of an STM exhibits an impedance $Z_{\text{jun}}$ of $10^8$ to $10^9$ Ω. In contrast, the impedance of the microwave transmission line built from standard microwave components is $Z_{\text{ref}} = 50$ Ω. The interface of the tunnel junction to the microwave transmission line thus represents a massive change of impedance, which affects the degree of reflection of an incoming microwave at the junction. The latter is described by the complex-valued scattering parameter[17]:

$$S_{11,\text{jun}} = \frac{Z_{\text{jun}} - Z_{ref}}{Z_{\text{jun}} + Z_{ref}} \quad (1)$$

Accordingly, the STM junction will almost completely reflect incoming microwaves, *i.e.* $S_{11} \approx 1$. Notice that $S_{11}$ is frequency-dependent, since the complex impedance of the tunnel junction depends on the frequency $f$ of the microwave as well as on the conductance $G$ and capacitance $C$ of the junction[17]:

$$Z_{\text{jun}} = \frac{1}{Y_{\text{jun}}} = \frac{1}{G + i\, 2\pi\, f\, C} \quad (2)$$

where $Y_{\text{jun}}$ is the complex-valued admittance and $i = \sqrt{-1}$ is the imaginary unit. Setting well-defined values for $Z_{\text{ref}}$ and $f$ leaves $S_{11}$ dependent on $G$ and $C$ of the junction only. This dependency lies at the heart of microwave reflectance spectroscopy implemented on a scanning probe instrument[18,19], relying on the accurate quantitative determination of $C$ from measured $S_{11}$ data. Such quantitative determination requires accurate calibration, because $S_{11,\text{jun}}$ cannot be measured directly. Instead, the experimental measurement yields the reflection coefficient $S_{11,\text{m}} \neq S_{11,\text{jun}}$. Notice that the subscript m indicates that $S_{11}$ is measured at the so-called measurement plane. $S_{11,\text{jun}}$ refers to the value of $S_{11}$ at the tunnel junction, also known as the calibration plane. The measurement plane is separated from the calibration plane by various circuit components and the transmission line. The amplitude and phase of the complex-valued $S_{11,\text{m}}$ systematically deviate from $S_{11,\text{jun}}$ due to microwave absorptions and reflections in the transmission line. Even though the deviation of $S_{11,\text{m}}$ from $S_{11,\text{jun}}$ is systematic, due to the complex nature of the transmission line it is not practical to use an analytic model to describe the relationship between $S_{11,\text{m}}$ and $S_{11,\text{jun}}$. The well-known numerical 3-error-term model[10,20,21] relates $S_{11,\text{jun}}$ to $S_{11,\text{m}}$ via three complex-valued parameters $e_{00}$, $e_{01}$, and $e_{11}$:

$$S_{11,\text{m}} = e_{00} + e_{01} \frac{S_{11,\text{jun}}}{1 - e_{11}\, S_{11,\text{jun}}} \quad (3)$$

Importantly, the phase and amplitude of $S_{11}$ depend strongly on the exact frequency chosen. Therefore, the *e*-parameters have to be determined separately for every measurement frequency.

Quantitative $G$ and $C$ values can only be derived from $S_{11,\text{m}}$ if the values of the *e*-parameters are precisely known. The *e*-parameters can be obtained from an *in-situ* retraction curve calibration. Retraction curve calibration requires that both $S_{11,\text{m}}(z)$ and $S_{11,\text{jun}}(z)$ are known [$z$ is the tip-sample separation, see Fig. 1(a)]. In an early approach by Farina *et. al.*[21], $C(z)$ was calculated using an assumed analytic function with absolute tip-sample distance and *ex-situ* measured tip-radius as input parameters. In AFM, a recent retraction calibration procedure relies on simultaneously measuring $\partial C(z)/\partial z$ and $S_{11,\text{m}}(z)$ during tip retraction[10]. Using Eqs. (1) and (2) and the assumption that $G = 0$, $S_{11,\text{jun}}(z)$ is calculated from the independently measured $C(z)$. Inserting $S_{11,\text{jun}}(z)$ and $S_{11,\text{m}}(z)$ into Eq. (3) allows to determine the three *e*-parameters via a least-squares fit routine. In STM, the force detection required to measure $\partial C(z)/\partial z$ is not available and an alternate strategy should be used.

In this work, we present a retraction calibration procedure that uses simultaneous $G(z)$ and $S_{11}(z)$ measurements upon the well-controlled variation of the tunnel distance $z$. We determine the numerical values of the *e*-parameters of Eq. (3) that define the calibration. Additionally, we present our implementation of microwave reflection spectroscopy in a low-temperature ultra-high vacuum STM. Our STM instrument has been upgraded by a cryogenic microwave transmission line[20,22]. The microwave circuit design described herein incorporates a custom interferometer and lock-in detection, forgoing the use of a vector network analyzer.

**Methods**

Microwave reflectance measurements were performed on our upgraded Createc low-temperature ultra-high vacuum STM system ($p < 4 \cdot 10^{-11}$ mbar, $T < 8.5$ K)[22,23], shown schematically in Fig. 1(a). All experiments herein were performed over a clean Au(111) surface (Surface Preparation Laboratory), prepared by repeated cycles of Ar ion sputtering (0.61 kV, 10 min) and annealing (703 K, 10 min). A tungsten tip (Bruker DTT10) was used as the STM and microwave probe. The bias voltage $V_{DC}$ was applied to the tip. The microwave signal was added to $V_{DC}$ via a bias-tee (SHF BT45R-B). Inside the vacuum chamber, the microwaves are transmitted via microwave compatible cables (Elspec MK5001 and Elspec Stormflex 047Cryo). At the sample side, a second bias-tee (Tektronix PSPL5541A) separates the DC line, which is grounded at the tunnel current amplifier (Femto DLPCA 200), and the microwave line, which is terminated with a 50 Ω load. Note that, as shown in Fig 1(a), the bias-tees are the only microwave related components inserted in the DC line, which does increase the tunnel current noise by approximately a factor of $< 2$. As the rest of the microwave circuit is fully separated from the DC line, this does not affect current noise or the signal-to-noise ratio of $G$ spectra.

We detect amplitude changes and phase shifts of the microwaves that have been reflected at the tunnel junction. In order to increase the signal-to-noise ratio, the background resulting from the near full microwave reflection at the tunnel junction [see Eq. (1)] was subtracted using a Mach-Zehnder type interferometer[11]. The orange shaded section of the schematic in Fig. 1(b) depicts the components and layout of the interferometer. Table I lists the details of the indicated components. The transmission line to the STM tip acts as the measurement signal of the interferometer [see pink arrows in Fig. 1(b)], while the variable voltage attenuator (VVA) acts as the difference signal (see cyan arrows in Fig. 1(b)]. Fig. 2(a) shows a schematic representation of the interferometer, including an indication of the microwave amplitude, $A$, and phase, $\varphi$, at different parts of the interferometer. An optimum (*i.e.* highest) signal-to-noise ratio for $S_{11,m}$ detection is achieved for destructive interference between the two paths in the interferometer. This happens when the signals from the two parts have the same amplitude ($A_3 = A_2$) and are antiphase ($\varphi_2 = \varphi_3 \pm (2n + 1)\pi$, where $n$ is an integer). Amplitude matching is performed by tuning the attenuation of the difference signal $A_3$ by the VVA input voltage. Antiphase conditions are obtained at fixed frequencies where the total path length of the two arms differ by exactly $(n + 1/2)\lambda$, where $n$ is an integer and $\lambda$ is the microwave wavelength. The frequencies for which this condition is satisfied are referred to as notches in the frequency spectrum[11], as shown in Fig. 2(b). Calibration measurements are performed at a frequency $f$ and VVA attenuation that maximizes destructive interference in the interferometer at the tunneling setpoint.

The green shaded section of the schematic in Fig. 1(b) depicts the down-modulation circuit and Table I lists the details of the indicated components. Before down-modulation, the microwaves coming from the interferometer are amplified using a low-noise amplifier (LNA). Then, the microwave signal is demodulated to an intermediate frequency $f_{IF}$, since the microwave frequency $f$ is too high to detect with the lock-in amplifier directly. The demodulated signal was detected using a lock-in amplifier, at a reference frequency of $f_{IF}$ obtained by demodulating a fraction of the microwaves going to the interferometer. With this arrangement, continuous phase-stable detection is achieved. In the experiments presented herein, $f = 0.908$ GHz and $f_{IF} = 500$ kHz. The peak-to-peak amplitude of the incident microwaves at the tunnel junction[22] was determined to be $V_{pp,MW} = 69$ mV. $G$ was measured by a lock-in amplifier (EG&G 5210) using a sinusoidal modulation of $V_{DC}$ ($V_{rms,lock-in} = 8$ mV, $f_{lock-in} = 775$ Hz).

Table I: Details of the microwave circuit components depicted in Fig. 1.

| Label | Component | Brand/type |
|---|---|---|
| DiC | Directional Coupler | Marki C13-0126 |
| HyC | Hybrid Coupler | Marki QH0R518 |
| PD | Power Divider | Marki PD0126 |
| M | 3-Port Mixer | Mini Circuits ZX05-63LH-S+ |
| LNA | Low-Noise Amplifier | Low Noise Factory LNR1_15A_SV |
| VVA | Variable Voltage Attenuator | Universal Microwave Corp. AG-U000-60V |
| Gen 1 | Generator 1 | Keysight E8257D |
| Gen 2 | Generator 2 | Rohde & Schwarz SMA100A |
| LIA2 | Lock-In Amplifier | Zurich Instruments MFLI (500 kHz) |

**Results**

Fig. 1(a) outlines the schematic of our retraction curve measurement set-up for the calibration of $S_{11}$ based on the simultaneous measurement of $G(z)$ and $S_{11,m}(z)$ during tip retraction. Fig. 3(a) shows $G(z)$ recorded while retracting the tip a distance of $\Delta z = 51$ nm away from the Au(111) surface. Prior to recording this retraction curve, the tip was approached by $\Delta z = -1$ nm from the tunneling setpoint (51 pA and 20 mV), representing $z = 0$, to the starting point of the retraction curve at $z = -1$ nm. This was done in order to achieve a sufficiently large change of $S_{11,m}(z)$, which is found to occur at a conductance close to atomic point contact, $1\,G_0$, while retaining an approximately exponential $G(z)$ relation in the tunneling regime, as shown in the lower panel of Fig. 3(a). Here, $G_0 = 2\,e^2/h \approx 77.48$ µS is the conductance quantum, $e \approx +1.6022\,10^{-19}$ C is the elementary charge and $h \approx 6.6261\,10^{-34}$ Js is the Planck constant. The red curves in Fig. 3(a) represent a numerical fit to the $G(z)$ curve with an exponential decay function, as is well-established for tunneling in STM[13]

$$G(z) = G_0\,e^{-2\kappa_{eff}(z-z_0)} \tag{4}$$

where $\kappa_{eff}$ and $z_0$ are parameters of the fit. From the fit, we find $\kappa_{eff} = 3.96 \pm 0.01$ nm$^{-1}$, which is significantly lower than the value of $\kappa \approx 11.9$ nm$^{-1}$ expected from the work function of Au(111) of $\phi \approx 5.4$ eV alone[13]. We remark that our $\kappa_{eff}$ value is consistent with literature values for similarly small tunnel distances. In particular, originating from different effects, such as a non-trapezoidal shape of the tunneling barrier[24], finite input impedance of the tunneling current amplifier[25], and the setting-in of bond formation between tip and sample close to point contact.

In our concurrent microwave reflection measurement, the amplitude of $S_{11,m}(z)$ decreases approximately linearly over the entire retraction range, as shown in the upper panel of Fig. 3(b). In contrast, the phase of $S_{11,m}(z)$, shown in the lower panel of Fig. 3(b), increases sharply in the first nm above the surface, followed by a nearly linear increase over the rest of the retraction curve. Fig. 3(c) shows the same $S_{11,m}(z)$ data in a different representation, *i.e.* plotted in the complex plane and the $z$-dependence shown as color gradient. Based on the shape of the $S_{11,m}(z)$ curve, we distinguish two different regimes, labeled high $G$ (red marker) and $G \approx 0$ (cyan marker). The high $G$ regime ranges from $z$ = -1 nm to $z \approx$ -0.8 nm and is characterized by a conductance of about $G > G_0 /3$. The $G \approx 0$ regime starts at $z \approx$ 4 nm.

In the next step, the values of the parameters $e_{00}$, $e_{01}$, and $e_{11}$ are determined from the experimental $S_{11,\mathrm{m}}(z)$ data with the help of a numerical least-squares fit ($R^2 \geq 0.91$) using Eq. (3). We obtain the values $e_{00} = (-1.7 \pm 0.4) + (0.0 \pm 0.1)i$, $e_{01} = (0.41 \pm 0.08) + (-2.3 \pm 0.4)i$, and $e_{11} = (0.91 \pm 0.02) + (0.79 \pm 0.07)i$. Here, the numerical error intervals of the *e*-parameters are obtained from numerical fitting (see also supplementary material Fig. 1). The value of $e_{00}$ depends on the constant $C$ in the calculation of $S_{11,\mathrm{jun}}(z)$, as it compensates for the offset between $S_{11,\mathrm{m}}(z)$ and $S_{11,\mathrm{jun}}(z)$. Possible interdependencies between $e_{01}$ and $e_{11}$ can be analysed from the mathematical form of Eq. (3). Since all quantities are complex valued, simple interdependencies are obtained only in certain cases. For instance, in the limit of $|S_{11,\mathrm{jun}}| \to \pm\infty$, Eq. (3) reduces to $S_{11,\mathrm{m}} = e_{00} - \frac{e_{01}}{e_{11}}$, revealing a trivial interrelation between $e_{01}$ and $e_{11}$. For the general case both $e_{01}$ and $e_{11}$ are required for a correct description. Notice that the numerical fit procedure concerns the difference between the experimental data (Fig. 3c) and the respective values predicted by Eq. (3), as follows: The scattering parameter at the tunnel junction $S_{11,\mathrm{jun}}(z)$ is calculated using Eqs. (2) and (1). The $G(z)$ measured by the lock-in amplifier is the starting point in the calculation of $S_{11,\mathrm{jun}}(z)$. The exponential fit to $G(z)$ of Eq. (4), as shown in Fig. 3(a), is substituted in for the conductance term in Eq. (2). For the unknown capacitance term, we apply a Taylor series approximation of $C(z) \approx C(d) + C'(d)(z - d) + \cdots$. Since we perform the numerical fitting in a narrow *z*-range of 0.54 nm, we consider only the constant and linear terms of the series and neglect higher terms. Notice that in the high $G$ regime, the exponential decay of $G(z)$ dominates the linear $C'(d)(z - d)$ term in the Taylor expansion of $C(z)$. Therefore, we also neglect the linear term and assume $C(z)$ = const. in Eq. (2) for the rest of our analysis. We remark that after successful calibration the validity of the assumption is confirmed, *i.e.* the *z*-range of $z$ = -1 to -0.46 nm used for calibration is sufficiently narrow (see discussion).

Fig. 4(a) shows our result of the calibration, where the experimental $S_{11,\mathrm{m}}(z)$ data from Fig. 3(c) has been transformed into calibrated admittance $Y_{\mathrm{jun}}(z)$ using the values of the *e*-parameters obtained from the numerical fit procedure above. The entire measured *z*-range between $z$ = -1 to 50 nm is calibrated this way. Notice that according to Eq. (2), $Y_{\mathrm{jun}}(z) = G + i\, 2\pi f\, C$ and thus the real and imaginary parts of $Y_{\mathrm{jun}}(z)$ correspond to *G* and *C*, respectively. After calibration, the different conduction regimes can still be clearly distinguished [compare Figs. 3(c) and 4(a)]. As expected, the high $G$ regime (red, $z <$ -0.8 nm) is dominated by a change of $G$, while $C$ remains almost constant. Conversely, the low-$G$ regime (cyan, $z >$ 4 nm) is dominated by a significant change of $C$, while the value of G is almost constant at $G \approx 0$.

Figs. 4(b) and (c) show the calibrated data of Fig. 4(a) in a different representation, *i.e* $G^{\mathrm{MW}}(z)$ and $\Delta C^{\mathrm{MW}}(z)$ respectively. To avoid confusion with the conventionally obtained $G$ data, the superscript MW indicates that the respective data has been obtained from the microwave reflection procedure. Note that a capacitance shift $\Delta C^{\mathrm{MW}}(z)$ is measured, since the magnitude of the capacitance between the sample and the tip-holder (stray capacitance) is not exactly known[10].

**Discussion**

The quality of the obtained calibration can be judged by comparing the conductance data, as shown in Fig. 4(b), that has been obtained in two different manners. Firstly, the exponential fit to $G(z)$, directly measured by the common $dI/dV$ lock-in method [see also Fig. 3(a)], which was used to obtain the *e*-parameters, is shown as the red curve. Secondly, $G^{MW}(z)$ obtained from the microwave reflection according to the procedure described above is shown as the blue curve. As can be seen, both curves are in reasonably good agreement. Notice that the agreement is particularly good in the calibration interval of $z < -0.46$ nm, as illustrated in the lower panel of Fig. 4(b). From this good quantitative agreement, we conclude that the calibration was successful. We remark that the minor discrepancy between the fit to $G(z)$ (red) and the $G^{MW}(z)$ data (blue) in the calibration interval cannot be explained by deviations in $C$ due to the tip apex. Instead, it may be rationalized by a frequency dependence of $G^{26,27}$ that is not explicitly included in the calibration procedure and goes beyond the scope of the present work.

In order to verify the validity of our assumption that the change in capacitance is negligible on the calibration interval of $z = -1$ to $-0.46$ nm, we compare the relative impact of changes in $G$ and $C$ on $Y_{\text{jun}}(z)$. We obtain a respective change of $\Delta G(z) = \Delta Re[Y_{\text{jun}}(z)] \approx 1.6\ G_0 \approx 126$ µS, see lower panel of Fig. 4(b). In comparison, a linear fit over the whole *z*-range from -1 to +50 nm of the $\Delta C^{MW}(z)$ curve, see Fig. 4(c), yields a value for the slope $\partial C^{MW}(z)/\partial z$ of -2.3 fF/nm, resulting in an absolute change of $\Delta C^{MW} \approx 1.2$ fF over the respective *z*-range ($\Delta z = 1 - 0.46 = 0.54$ nm), and finally resulting in a value of $\Delta C^{MW} 2\pi f = \Delta \text{Im}[Y_{\text{jun}}(z)] \approx 7$ µS. This value is circa 20 times smaller (!) than the above value of 126 µS, demonstrating that $Y_{\text{jun}}$ is indeed dominated by $G$ (and not by $C$), in agreement with the above assumption. Notice that within the calibration interval the slope $\partial G(z)/\partial z$ derived from Fig. 4(b) is found to be always larger than the respective slope of $2\pi f\ \partial C^{MW}(z)/\partial z$ (see above), which is again in agreement with the above assumption. Repeating the calibration for slightly larger and smaller *z*-ranges ($\pm$ 0.05 nm) yields values that are in good agreement [see supplementary material Fig. 2 for deviations in $G^{MW}(z)$ and $\Delta C^{MW}(z)$] to the above values. Either enforcing the $G = 0$ condition for $z > 10$ nm or accounting for the linear term in $C(z)$ in a self-consistent optimization loop, again yields consistent values. This shows that the calibration procedure yields robust values for the capacitance.

The *e*-parameters obtained in this calibration procedure numerically describe non-idealities, such as absorptions and reflections, of microwave propagation through the transmission line from the calibration plane to the measurement plane. As such, these *e*-parameters are independent from *z* and from tip and sample properties[10]. The calculation of $S_{11,\text{jun}}(z)$ involves an exponential fit to $G(z)$, so the tip must be stable to give an exponential $G(z)$ dependence during calibration. The *e*-parameters can be used to calibrate microwave reflectance measurements to obtain quantitative $\Delta C^{MW}$ values, as shown in Fig. 4(c). It is important to note that a particular set of *e*-parameters is only valid for the frequency and VVA settings at which the calibration was performed. To obtain a frequency spectrum, a calibration should be performed for each frequency. Currently, this is done with a separate approach curve at each frequency notch, but we envision using a full frequency sweep during a single approach curve in the future.

**Conclusion**

We present a calibration procedure for microwave reflectance spectroscopy based on the acquisition of conductance retraction curves in an ultra-high vacuum low-temperature STM instrument. While the retraction curve calibration concept based on capacitive force detection has been very successfully employed for microwave reflection measurements in an AFM[10] in the past, in STM the force detection is not available. Therefore, we acquire the tunneling conductance, which changes exponentially with distance, instead of the capacitance, which has a long-range interaction and changes only slowly with tip sample distance. We show that in the high $G$ regime, for approximately $G > G_0/3$ occurring at sub-nm tip-sample distances, the change of the measured scattering parameter $S_{11,\text{m}}$ with respect to the tunnel distance is dominated by changes in the conductance signal. In this narrow $z$-range of 0.54 nm in the high-conduction regime, the change of the junction capacitance with tunnel distance can be neglected. This allows for a successful calibration of the error parameters describing the distortion of $S_{11,\text{m}}$ relative to the scattering parameter at the tunnel junction $S_{11,\text{jun}}$. We remark that the employed $e$-parameter procedure captures all non-idealities of the transmission line and detection circuit and can be readily extended to calibration of an entire range of frequencies.

The calibration of microwave reflectance spectra presented here will pave the way for quantitative capacitance and dielectric spectroscopy measurements with a lateral resolution of a few nm and ultimately single molecule fingerprinting in low-temperature ultra-high vacuum STM. Specifically, we expect that the tip-sample contribution to $\Delta C^{\text{MW}}$ can be isolated either in constant height measurements or with an appropriate capacitance model. This may allow us to determine material properties, such as the dielectric permittivity, in future experiments[12]. The imaging capabilities of STM together with frequency dependent microwave reflectance, could then be utilized to achieve full spectroscopic imaging.

**Supplementary material**

See the supplementary material for additional details on the dependence of the $e$-parameters, $G^{\text{MW}}(z)$, and $\Delta C^{\text{MW}}(z)$ on the $z$-range included in the calibration procedure.

**Acknowledgments**

The authors acknowledge the financial support the European Research Council (ERC) under the European Union's Horizon 2020 research and innovation programme (Grant Agreement No. 771193). This work was additionally supported by funding from EMPIR project 20IND12 Elena.

**Author declarations**

**Conflict of Interest**

The authors have no conflicts to disclose.

**Author Contributions**

Bareld Wit: Conceptualization, Formal Analysis, Methodology, Validation, Visualization, and Writing. Georg Gramse: Conceptualization, Funding Acquisition, Methodology, Validation, and Writing. Stefan Müllegger: Conceptualization, Validation, Funding Acquisition, Project Administration, Supervision, and Writing.

**Data availability**

The data that support the findings of this study are available from the corresponding author upon reasonable request.

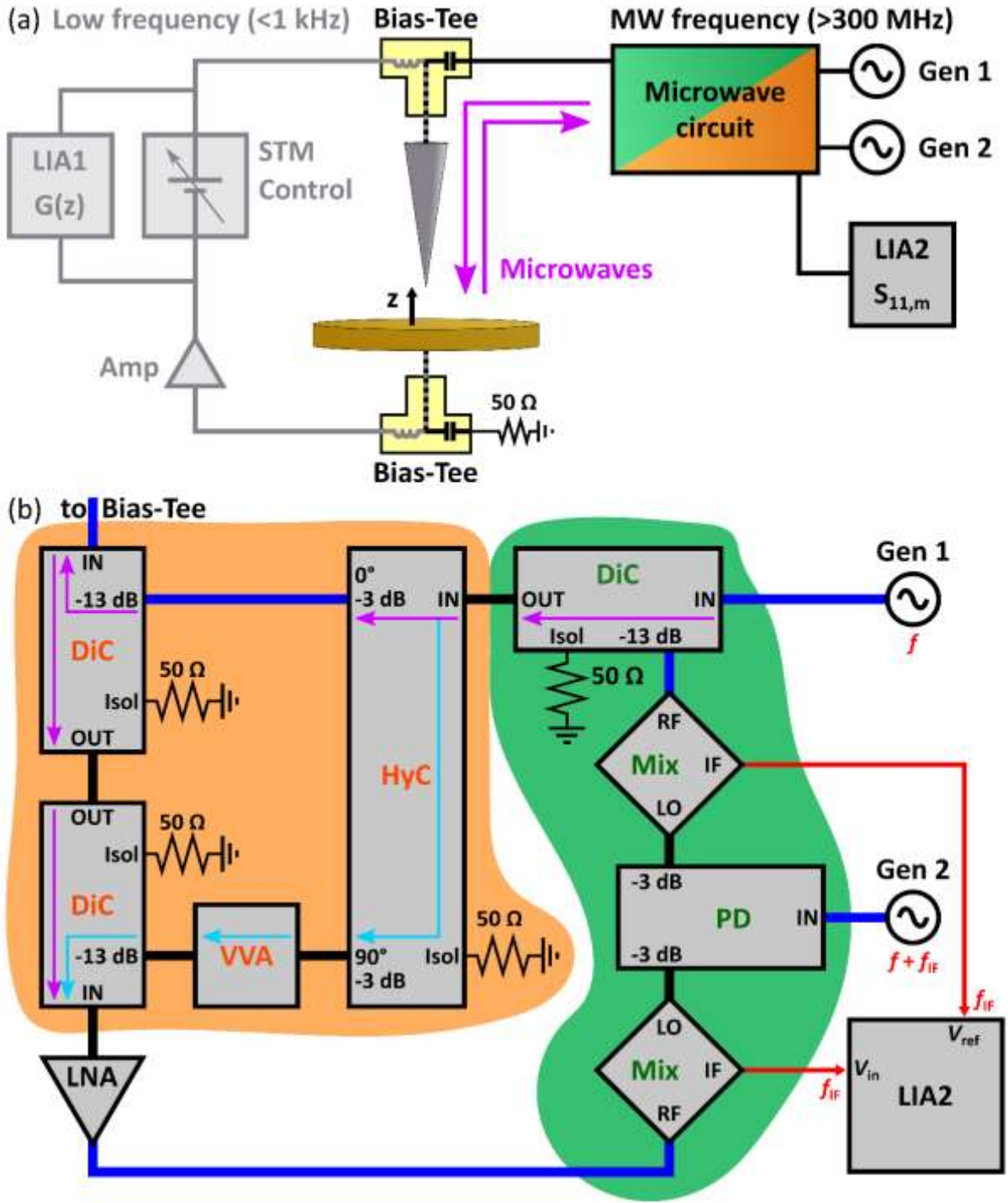

FIG. 1: The experimental set-up. (a) Schematic overview of the electronic circuit. Low frequency components for STM operation and $G$ detection are outlined in grey. Microwave specific components are outlined in black. The microwave voltage is added to $V_{DC}$ using a bias-tee (colored yellow). (b) Block diagram of the microwave interferometer circuit (shaded orange) and the down-conversion circuit (shaded green). Pink and cyan arrows indicate the measurement and difference paths of the microwaves through the interferometer, respectively. Specific ports of the components are labeled. Black lines indicate connections via a single adapter piece, blue lines are dedicated microwave cables, and red lines are low-noise BNC cables. For abbreviations and details of the components, see Table I.

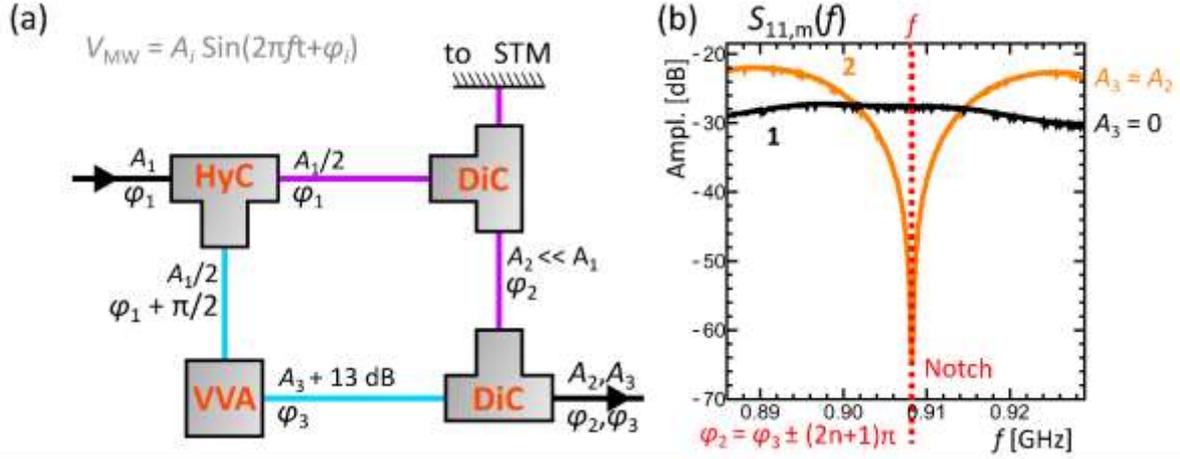

FIG. 2: Details of the interferometer. (a) Schematic representation of the interferometer. The purple and cyan lines represent the measurement and difference paths of the microwaves through the interferometer, as in Fig. 1(b). Amplitudes $A_i$ and phases $\varphi_i$ of $V_{MW}$ are indicated at different points in the interferometer. In the directional coupler where the two paths meet, the two waves interfere and the resulting $V_{MW}$ depends on the amplitudes and phases of the interfering waves. (b) Microwave scattering parameter $S_{11,m}(f)$ amplitude measured without (black curve 1) and with (orange curve 2) the interferometer. Curve 1 is obtained with $A_3 = 0$, so no interference takes place and the output amplitude is solely determined by the loss in the microwave circuit. For curve 2, $A_3$ is tuned to match $A_2$. The relative phase of the interfering waves is determined by the frequency $f$. When $f$ is such that the two waves are out of phase, *i.e.* $\varphi_2 = \varphi_3 \pm (2n+1)\pi$, destructive interference is maximized, as indicated with a red dashed line.

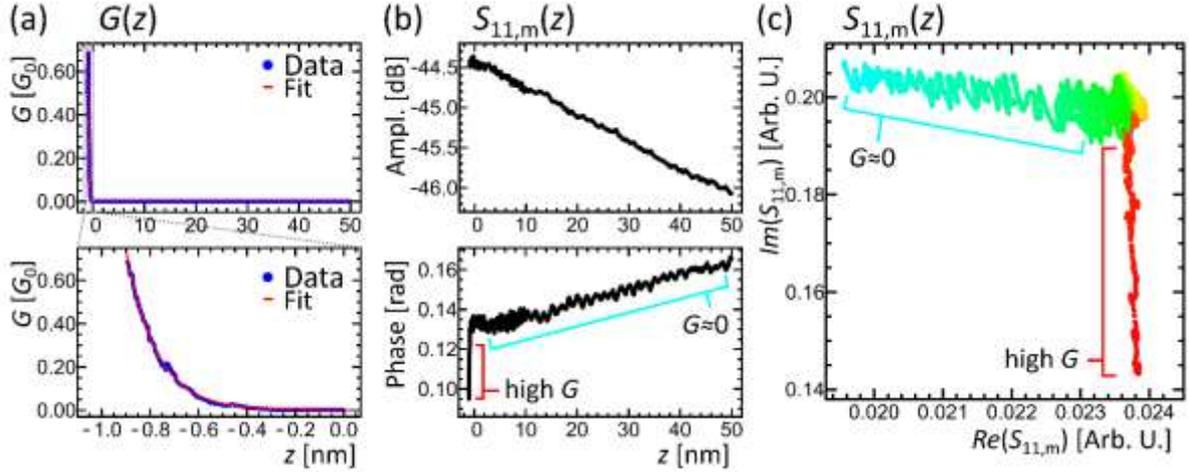

FIG. 3: Experimental retraction curves. (a) Conductance retraction curve $G(z)$ recorded over clean Au(111). The lower panel shows the portion of $G(z)$ at $z < 0$. The red curves represent a fit with an exponential decay function ($R^2 = 0.9986$). (b) $z$-dependence of amplitude (upper panel) and phase (lower panel) of the complex $S_{11,m}(z)$ during the microwave reflection retraction curves. (c) $S_{11,m}$ plotted in the complex plane. The color code is non-linear with respect to $z$, with red being points taken closest to the surface and green/cyan furthest away from the surface. All $z$-values are reported relative to the tunneling setpoint, with negative numbers meaning closer to the surface.

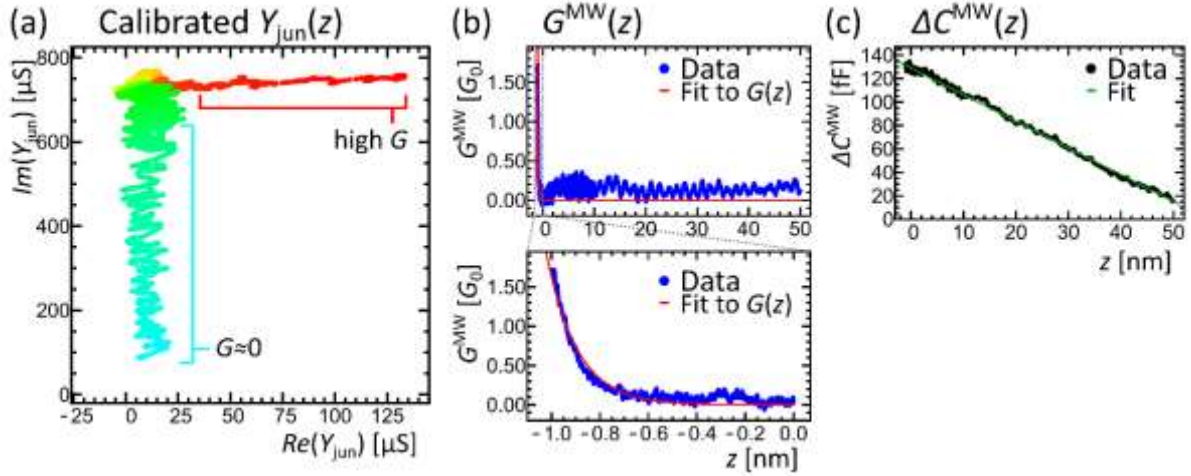

FIG. 4: The result of calibration. (a) Calibrated admittance $Y_{\text{jun}}$ (reciprocal impedance) plotted in the complex plane. (b) The conductance $G^{\text{MW}}(z) = Re[Y_{\text{jun}}(z)]$ derived from $S_{11,\text{jun}}(z)$ after calibration. The lower panel shows the portion of $G^{\text{MW}}(z)$ at $z < 0$. The red curves represent the fit to $G(z)$ as measured with the conventional lock-in method. These curves are included to directly compare $G(z)$ with $G^{\text{MW}}(z)$; in a successful calibration the conductance curves coincide. (c) The capacitance change $\Delta C^{\text{MW}}(z) = Im[Y_{\text{jun}}(z)]/(2\pi f)$ derived from $S_{11,\text{jun}}(z)$ after calibration. The green line is a linear fit ($R^2 = 0.9957$) to $\Delta C^{\text{MW}}(z)$. All $z$-values are reported relative to the tunneling setpoint, with negative numbers meaning closer to the surface.